\documentclass[journal=langd5,manuscript=article]{achemso}
\usepackage[version=3]{mhchem}

\author{Alexandre P. dos Santos}
\affiliation[Universidade Federal do Rio Grande do Sul]
{Instituto de F\'isica, Universidade Federal do Rio Grande do Sul, Caixa Postal 15051, CEP 91501-970, Porto Alegre, RS, Brazil}
\author{Alexandre Diehl}
\affiliation[Universidade Federal de Pelotas]
{Departamento de F\'isica, Instituto de F\'isica e Matem\'atica, Universidade Federal de Pelotas, Caixa Postal 354, CEP 96010-900, Pelotas, RS, Brazil}
\author{Yan Levin}
\email{levin@if.ufrgs.br}
\affiliation[Universidade Federal do Rio Grande do Sul]
{Instituto de F\'isica, Universidade Federal do Rio Grande do Sul, Caixa Postal 15051, CEP 91501-970, Porto Alegre, RS, Brazil}

\title{Surface tensions, surface potentials and the Hofmeister series of electrolyte solutions}

\begin{document}

\begin{abstract}

A theory is presented which allows us to accurately calculate the surface tensions and the surface potentials of electrolyte solutions. Both the ionic hydration and the polarizability are taken into account. We find a good correlation between the Jones-Dole viscosity $B$-coefficient and the ionic hydration near the air-water interface. The kosmotropic anions such as fluoride, iodate, sulfate and carbonate, are found to be strongly hydrated and are repelled from the interface. The chaotropic anions such as perchlorate, iodide, chlorate and bromide are found to be significantly adsorbed to the interface. 
Chloride and bromate anions become weakly hydrated in the interfacial region.  The sequence of surface tensions and surface potentials is found to follow the Hofmeister ordering. The theory, with only one adjustable parameter, quantitatively accounts for the surface tensions of 10 sodium salts for which there is experimental data.
\end{abstract}

\section{Introduction}

Electrolyte solutions have been the subject of intense study for over a century. While the bulk properties of electrolytes are now quite well understood, their behavior at interfaces and surfaces still remains a puzzle. Over a hundred years ago, Hofmeister observed that presence of electrolyte in water modified significantly solubility of proteins. While some salts lead to protein precipitation (salting out), other salts stabilize proteins increasing their solubility (salting in). A few years after this curious observation, Heydweiller \cite{He10} discovered that salt dissolved in water increased the surface tension of  solution-air interface. While cations had only a small influence on the surface tension, anions affected it quite significantly. Furthermore, the magnitude of the variation of the surface tension followed the same sequence discovered by Hofmeister earlier. It appeared that the two effects were related.

Because of its great importance for biology, over the last century there has been a tremendous effort to understand the ionic specificity. Since air-water interface is relative simple, as compared to proteins, most of the theoretical work has concentrated on it. Langmuir \cite{La17} was the first to attempt a theoretical explanation of the physical mechanism behind the increase of the surface tension produced by electrolytes. Using the Gibbs adsorption isotherm equation, Langmuir concluded that the phenomenon was a consequence of ion depletion near the air-water interface, and suggested that the depleted
layer was about $4$ \AA\ in width. No clear explanation for existence of this ion-free  layer was provided by Langmuir,
and soon it became clear that in order to obtain
a reasonable agreement with experiments, its width had to be a function of ionic concentration \cite{HaMc25,*HaGi26}. A further insight was provided by Wagner \cite{Wa24}, based on the Debye-H\"uckel (DH) theory of strong electrolytes \cite{DeHu23}. Wagner argued that ionic depletion was a consequence of the electrostatic repulsion produced by the interaction of ions with their electrostatic images across the air-water interface. Wagner's theory was quite complicated and a simplified version was proposed by Onsager and Samaras (OS), who also derived a limiting law which, they argued, had a universal validity for all electrolytes at sufficiently low concentrations \cite{OnSa34}. Indeed, careful experimental measurements have confirmed the OS limiting law \cite{LoNu42,*Pa59}. However at larger concentrations, OS theory was found to strongly underestimate surface tensions.

Similar to  Langmuir, Wagner and OS integrated the Gibbs adsorption isotherm equation to obtain the excess surface tension. Some time ago, Levin and Flores-Mena (LFM) proposed a different approach based on the direct Helmholtz free energy  calculation \cite{LeMe01}. The LFM theory combined Langmuir, Wagner, and OS insights into one theory. They argued that in addition to the ion-image interaction, ionic hydration leads to a hard-core-like repulsion from the Gibbs dividing surface. Using a $2$ \AA\ hydrated radius of \ce{Na+} and \ce{Cl-}, they were able to obtain a very good agreement with the experimental measurements of surface tension of \ce{NaCl} solution, up to $1$ M concentration. However, the LFM theory failed to correctly account for the ionic specificity, predicting that the surface tension of \ce{NaI} salt should be larger than that of \ce{NaCl}, contrary to experiments. It was clear that the LFM theory was still lacking an important ingredient. An indication of the missing ingredient was already present in the work of Frumkin, 80 years earlier \cite{Fr24}. Frumkin measured the electrostatic potential difference across the solution-air interface and found that for all halogen salts, except fluoride, potential was lower in air than in water. This meant that anions were preferentially solvated near the interface. Bostr\"om et al. \cite{BoWi01} suggested that this ionic specificity, was a consequence of dispersion forces arising from finite frequency electromagnetic fluctuations.  Their theory, however, predicted surface potentials of opposite sign to the ones measured by Frumkin, implying that cations were preferentially adsorbed to the interface.  This was clearly contradicted by the 
simulations on small water clusters \cite{PeBe91,*DaSm93,*StBr99}, as well as by the subsequent large scale polarizable force fields simulations \cite{JuTo02,*JuTo06,*HoNe07,*Br08} and by the photoelectron emission experiments \cite{MaPo91,*Gh05,*Ga04}.  All these agreed with Frumkin that
large anions, and not small cations, that are preferentially solvated near the air-water interface. 
A theoretical explanation for this behavior was advanced by Levin \cite{Le09}, who argued that anionic surface solvation was a consequence of 
the competition between the 
cavitational and the electrostatic energies.   The cavitational energy arises from the perturbation to the hydrogen bond network produced
by ionic solvation. This results in a short range force which drives ions towards the interface.  This is counterbalanced by the electrostatic Born solvation force which arises from the dipolar screening of the ionic self-energy in aqueous environment.  For hard (weakly polarizable) ions, the Born energy
is much larger than the cavitational energy, favoring the bulk solvation.  However for large polarizable ions, the energy balance is shifted.
For such ions, ionic charge can easily redistribute itself so that even if a large fraction of ionic volume is exposed to air, the electrostatic energy
penalty for this remains small, since most of the ionic charge remains hydrated. This means that through surface solvation, large, strongly polarizable ions, can have the best of two worlds --- gain the cavitational energy at a small price in electrostatic self energy.  
Levin derived the interaction potential quantifying this effect \cite{Le09}. In a follow up paper, Levin et al. \cite{LeDo09}
used this potential to quantitatively account for the surface tensions of all sodium-halide salts. In this paper we will extend the theory of reference \cite{LeDo09} to calculate the surface tensions and the surface potentials of others sodium salts and to derive the Hofmeister series.

\section{Model and theory}

Consider an electrolyte solution confined to a mesoscopic drop of water of radius $R$, which corresponds to the position of the Gibbs dividing surface (GDS) \cite{HoTs03,LeDo09}. We define the adsorption (ion excess per unit area) as
\begin{equation}
\label{e1}
\Gamma_\pm=\frac{1}{4 \pi R^2} \left[\int_0^\infty \rho_\pm(r) 4 \pi r^2 {\rm d}r -\frac{4 \pi R^3}{3} c_b \right] \ ,
\end{equation}
where $\rho_\pm(r)$ are the ionic density profiles and $c_b=\rho_+(0)=\rho_-(0)$ is the bulk concentration of electrolyte. If $N$ ion pairs are inside the drop, eq \ref{e1} simplifies to $\Gamma\pm=N/4\pi R^2 - c_b R/3$.

\begin{figure}
\begin{center}
\includegraphics[width=7.29cm]{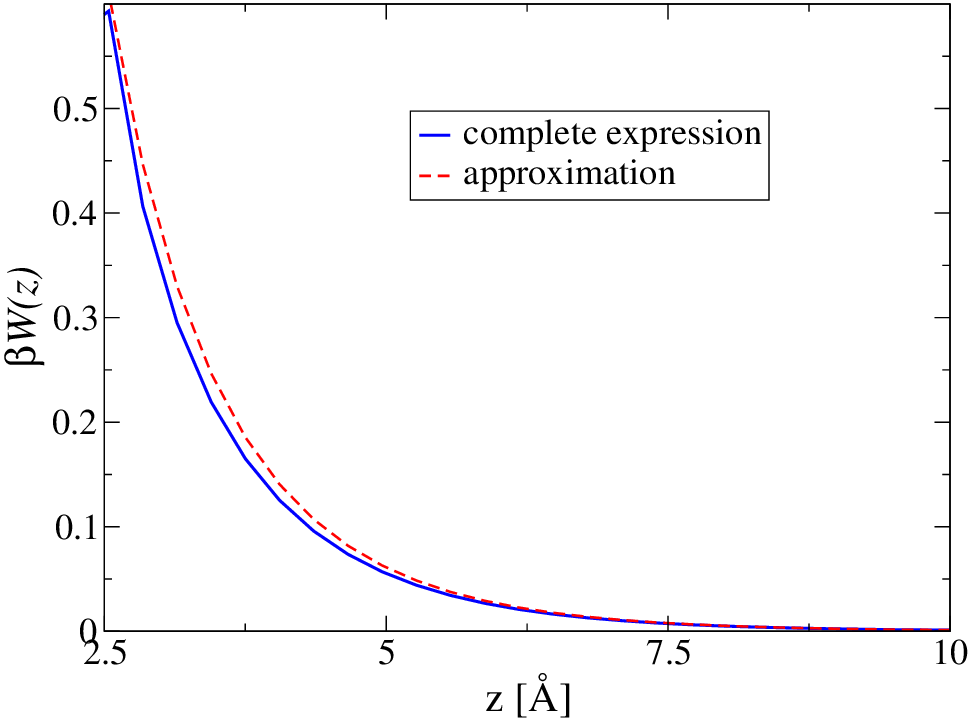}
\end{center}
\caption{Comparison between the complete expression (eq \ref{e2}) and its approximation (eq \ref{e2ap}), for an ion of radius $2.5$ \AA at 1M concentration.}
\label{fig1}
\end{figure}

The water and air will be treated as uniform dielectrics of permittivities $\epsilon_w=80$ and $\epsilon_a=1$, respectively. The surface tension can be obtained by integrating the Gibbs adsorption isotherm equation, $ {\rm d} \gamma=-\Gamma_+ {\rm d} \mu_+ - \Gamma_- {\rm d} \mu_-$, where  $\beta \mu_\pm=\ln(c_b \Lambda_\pm^3)$ are the  chemical potentials and $\Lambda_\pm$ are the de Broglie thermal wavelengths. Let us first consider alkali-metal cations, such as lithium, sodium, or potassium.  Because these cations are small, they have large surface charge density, which leads to strong interaction with surrounding water molecules, resulting in an effective hydrated  radius $a_h$.  We can, therefore, model these ions as hard spheres of radius $a_h$ with a point charge $q$ located at the origin. Because of their strong hydration, these cations can not move across the GDS since this would require them to shed their solvation sheath.  For mesoscopic drops we can neglect the curvature of the GDS. To bring a cation from bulk electrolyte to some distance $z>a_h$ from the GDS then requires  \cite{LeMe01,LeDo09}
\begin{equation}
\label{e2}
W(z;a_h)= \frac{q^2}{2\epsilon_w} \int_0^\infty dk e^{-2 s (z-a_h)} \frac{k[ s \cosh(k a_h)-k \sinh(k a_h)]}
{s[ s \cosh(k a_h)+ k \sinh(k a_h)]} \ ,
\end{equation}
of work.  The GDS is located at $z=0$ and the axis is oriented into the drop. We have defined $s=\sqrt{\kappa^2 + k^2}$, where $\kappa=\sqrt{8 \pi q^2 c_b/\epsilon_w k_B T}$ is the inverse Debye length. The eq \ref{e2} is well approximated by
\begin{equation}
\label{e2ap}
W_{ap}(z;a_h)=\frac{W(a_h;a_h) a_h}{z} \ e^{-2 \kappa (z-a_h)} \,,
\end{equation}
see \ref{fig1}. This form will be used later to speed up the numerical calculations.

Unlike small cation, large halogen anions of bare radius $a_0$ have low electronic charge density and are weakly hydrated. The polarizability of an anion is $\gamma_a$ and we define its relative polarizability  as $\alpha=\gamma_a/a_0^3$.  We can model these ions as imperfect spherical conductors.  When an anion is far from the interface, its charge $-q$ is uniformly distributed over its surface. However, when an anion begins to cross the GDS, its charge starts to redistribute itself on the surface so as to leave most of it in the high dielectric environment \cite{Le09}. The fraction of charge $x$ which remains hydrated when the ionic center is at distance $z$ from the GDS is determined by the minimization of the polarization energy \cite{Le09},
\begin{equation}\label{Upol}
U_p(z,x)= \frac{q^2}{2 a_0 \epsilon_w}\left[\frac{\pi x^2}{\theta(z)}+\frac{\pi [1-x]^2 \epsilon_w}{[\pi-\theta(z)]\epsilon_o}\right] + \frac{g}{\beta} \left[ x-\frac{1-cos[\theta(z)]}{2} \right]^2 \ ,
\end{equation}
where $\theta(z)=\arccos[-z/a_0]$ and $g=(1-\alpha)/\alpha$. We find
\begin{equation}
x(z)=\left[ \frac{\lambda_B \pi \epsilon_w}{a_0 \epsilon_o \left[\pi-\theta(z)\right]}+g [1-cos[\theta(z)]] \right] / \left[\frac{\lambda_B \pi}{a_0 \theta(z)} + \frac{\lambda_B \pi \epsilon_w}{a_0 \epsilon_0 [\pi-\theta(z)]} +2 g \right] \ .
\end{equation}
Substituting this back into eq \ref{Upol}, we obtain the polarization potential. This potential is repulsive, favoring ions to move towards the bulk. Nevertheless, the repulsion is quite soft compared to a hard-core-like repulsion of strongly hydrated cations. The force that drives anions toward the interface arises from cavitation. When ion is dissolved in water it creates a cavity from which water molecules are expelled. This leads to a perturbation to the hydrogen bond network and an energetic cost. For small voids, the cavitational energy scales with their volume. \cite{LuCh99,*Ch05}.  As an ion moves across the interface, the cavity that it creates in water diminishes proportionally to the fraction of the volume exposed to air \cite{Le09,LeDo09}, producing a short-range
interaction potential that forces ions to move across the GDS,
\begin{eqnarray}
\label{e3}
U_{cav}(z)=\left\{
\begin{array}{l}
 \nu a_0^3 \text{ for } z \ge  a_0  \ , \\
 \frac{1}{4} \nu a_0^3  \left(\frac{z}{a_0}+1\right)^2 \left(2-\frac{z}{a_0}\right) \text{ for } -a_0<z<a_0 \ ,
\end{array}
\right.
\end{eqnarray}
where $\nu \approx 0.3 k_B T/$ \AA$^3$ is obtained from bulk simulations \cite{RaTr05}.
For small strongly hydrated cations, this energy gain does not compensate for
the electrostatic energy penalty of exposing ionic charge to the low dielectric environment. 
For large polarizable ions, on the other hand, electrostatic energy penalty is small, and the cavitational
energy is sufficient to favor the surface solvation.  The total potential felt
by an unhydrated anion is then  \cite{LeDo09}
\begin{eqnarray}
\label{e5}
U_{tot}(z)=\left\{
\begin{array}{l}
W(z;a_0)+\nu a_0^3+\frac{q^2}{2 \epsilon_w a_0} \, \text{ for } z \ge  a_0 \ , \\
W(a_0;a_0) z/a_0 + U_p(z)+U_{cav}(z)\, \text{ for } 0<z<a_0 \ , \\
U_p(z)+U_{cav}(z)\, \text{ for } -a_0<z \le 0 \ .
\end{array}
\right.
\end{eqnarray}

The density profiles can now be calculated by integrating the non-linear modified Poisson-Boltzmann equation (mPB):
\begin{eqnarray}
\label{e7}
\nabla^2 \phi(r)&=&-\frac{4\pi q }{\epsilon_w} \left[\rho_+(r)-\rho_-(r)\right] \nonumber \ , \\
\rho_+(r)&=&\frac {N \Theta (R-a_h-r) e^{-\beta q \phi(r)-\beta W(z;a_h)}}{\int_0^{R-a_h} 4\pi r^2\, dr\, e^{-\beta q \phi(r)-\beta W(z;a_h)}} \ , \\
\rho_-(r)&=&\frac{N e^{\beta q \phi(r) -\beta U_{tot}(r)}}
{\int_0^{R+a_0} 4\pi r^2\, dr \, e^{\beta q \phi(r) -\beta U_{tot}(r)}} \nonumber \ ,
\end{eqnarray}
where $\Theta$ is the Heaviside step function. To speed up the numerical calculations we can replace $W \rightarrow W_{ap}$.

\section{Sodium-Halogen Salts}

First, we study the sodium-halogen salts \cite{LeDo09}. The anion radii were obtained by Latimer, Pitzer and Slansky \cite{LaPi39} by fitting the experimentally measured free energies of hydration to the Born model.  
Since our theory in the bulk also reduces to the Born model, these radii are
particularly appropriate: $a_I=2.26$ \AA,  $a_{Br}=2.05$ \AA,  $a_{Cl}=1.91$ \AA\ and  $a_{F}=1.46$ \AA. For ionic polarizabilities we use the values from the reference \cite{PyPi92}: $\gamma_I=7.4$ \AA$^3$, $\gamma_{Br}=5.07$ \AA$^3$, $\gamma_{Cl}=3.77$ \AA$^3$ and  $\gamma_{F}=1.31$ \AA$^3$.

\begin{figure}
\begin{center}
\includegraphics[width=7.29cm]{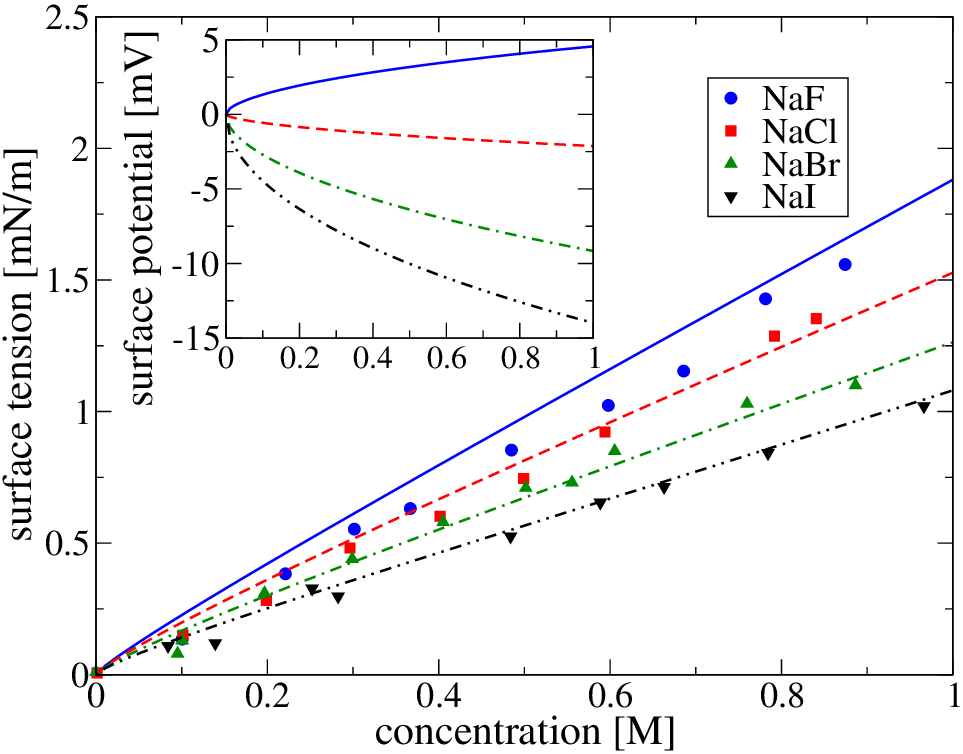}
\end{center}
\caption{Surface tensions for \ce{NaF}, \ce{NaCl}, \ce{NaBr} and \ce{NaI}. The symbols are the experimental data \cite{MaTs01,MaMa99,Ma} and the lines are the results of the present theory. The inset shows the surface potential difference as a function of molar concentration.}
\label{fig2}
\end{figure}
The excess surface tension can be obtained by integrating the Gibbs adsorption isotherm equation (eq \ref{e1}). We start with \ce{NaI}. Since \ce{I-} is large and soft, it should be unhydrated in the interfacial region. Adjusting the hydrated radius of \ce{Na+} to best fit the experimental data \cite{MaTs01} for \ce{NaI}, we obtain \cite{LeDo09} $a_{h}=2.5$ \AA. We will use this partially hydrated radius of \ce{Na+} in the rest of the paper. Considering that \ce{Br-} is also large and soft, we expect that it will also remain unhydrated in the interfacial region.  This expectation is well justified
and we obtain a very good agreement with the experimental data \cite{Ma}, see \ref{fig2}. For \ce{F-} the situation should be very 
different.  This ion is small, hard and strongly hydrated. This means that just like for a cation, a hard core repulsion 
from the GDS must be explicitly included in the mPB equation.  For hydrated (or partially hydrated) anions the density is then 
\begin{eqnarray}
\label{e7a}
\rho_-(r)&=&\frac {N \Theta (R-a_h-r) e^{\beta q \phi(r)-\beta W(z;a_h)}}{\int_0^{R-a_h} 4\pi r^2\, dr\, e^{\beta q \phi(r)-\beta W(z;a_h)}} \ ,
\end{eqnarray}
Using this in eq \ref{e7}, an almost perfect agreement with the experimental data \cite{MaTs01} is found for \ce{NaF}, using the usual bulk hydrated radius of \ce{F-}, $a_h=3.52$ \AA \cite{Ni59},  see \ref{fig2}. The table salt, \ce{NaCl}, is the most difficult case to study theoretically, 
since \ce{Cl-} is sufficiently small to 
remain partially hydrated near the GDS. We find a good fit to the experimental data \cite{MaMa99}, \ref{fig2}, 
using a partially hydrated radius of \ce{Cl-},  $a_h=2.0$ \AA, which is very close to its bare size.

\begin{figure}
\begin{center}
\includegraphics[width=7.29cm]{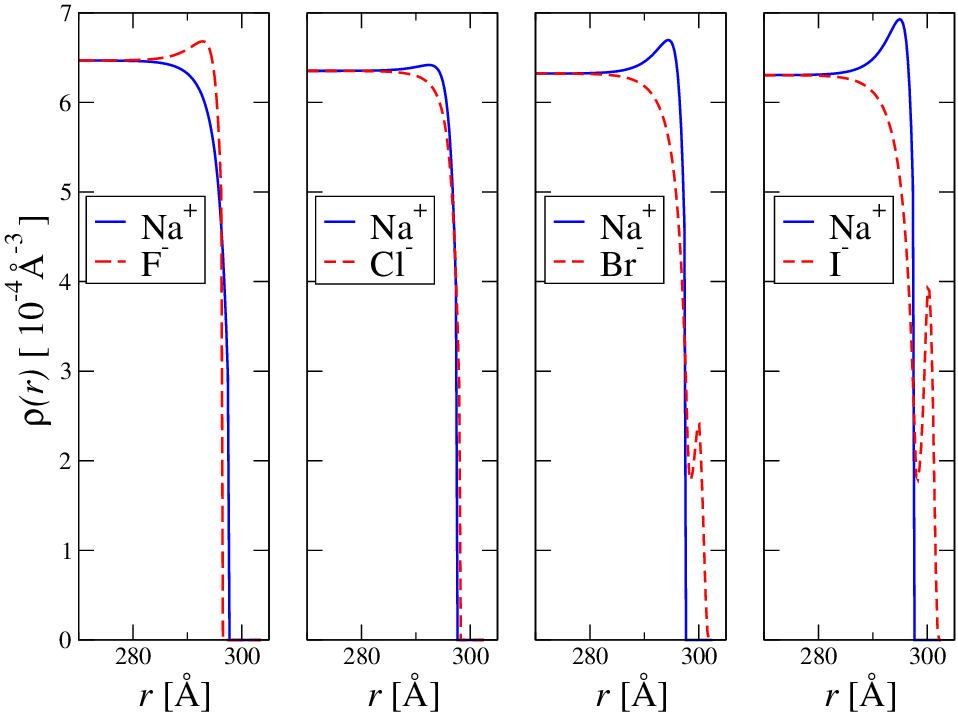}
\end{center}
\caption{Theoretical ionic density profiles for \ce{NaF}, \ce{NaCl}, \ce{NaBr} and \ce{NaI} at $1$M. The GDS is at $r=300$ \AA.}
\label{fig3}
\end{figure}
The ionic density profiles for these salts are plotted in \ref{fig3}. In agreement with the polarizable force fields simulations
\cite{PeBe91,*DaSm93,*StBr99,JuTo02,*JuTo06,*HoNe07,*Br08}, the large halogens \ce{I-} and \ce{Br-} are strongly adsorbed at the interface.
However, just as was found using the vibrational sum-frequency spectroscopy \cite{RaRi04}, their concentration at the GDS remains below that in the bulk. The electrostatic potential difference across the air-water interface is $\chi=\phi(\infty)-\phi(0)$.  We find that our theoretical values for the surface potentials are in reasonable agreement with the measurements of Frumkin \cite{Fr24,Ra63} and Jarvis et al. \cite{JaSc68}, \ref{tab1}.
\begin{table}
   \centering
   \setlength{\belowcaptionskip}{10pt}
   \caption{Surface potentials difference at $1$ M for various salts}
   \begin{tabular}{|c|c|c|c|}
      \hline
                   &    Calculated (mV)   &   Frumkin \cite{Fr24,Ra63} (mV)  &   Jarvis et al. \cite{JaSc68} (mV)  \\
      \hline
      \ce{NaF}     &      4.7             &        --       &         --           \\
      \hline
      \ce{NaCl}    &     -2.1             &        -1       &     $\approx$ -1     \\
      \hline
      \ce{NaBr}    &     -9.4             &        --       &     $\approx$ -5     \\
      \hline
      \ce{NaI}     &     -14.3            &       -39       &     $\approx$ -21    \\
      \hline
      \ce{NaIO3}   &      5               &       --        &       --             \\
      \hline
      \ce{NaBrO3}  &     -0.12            &       --        &       --             \\
      \hline
      \ce{NaNO3}   &      -8.27           &       -17       &     $\approx$  -8    \\
      \hline
      \ce{NaClO3}  &     -11.02           &       -41       &       --             \\
      \hline
      \ce{NaClO4}  &     -31.1            &       -57       &       --             \\
      \hline
      \ce{Na2CO3}  &      10.54           &         3       &     $\approx$   6    \\
      \hline
      \ce{Na2SO4}  &      10.17           &         3       &     $\approx$   35   \\
      \hline
   \end{tabular}
   \label{tab1}
\end{table}

\section{Oxy-Anion Salts}

We next study the surface potentials and the surface tensions of sodium salts with more complex anions, such as: chlorate, nitrate, bromate, iodate, perchlorate, sulfate and carbonate. Naively one might expect that because of their large size, 
these oxy-anions will loose their hydration near the GDS.  This, however, is not
necessarily the case.  At the moment there is no reliable theory of hydration, nevertheless, it has been known for a long time that ions can be divided into two categories: structure-makers (kosmotropes) and structure-breakers (chaotropes).
This dichotomy is reflected, for example, in the Jones-Dole viscosity $B$-coefficient.  It is found experimentally that for large dilutions, viscosity varies with the concentration of
electrolyte as \cite{JeMa95}:
\begin{equation}
\eta_r = 1 + A \rho^{1/2} + B \rho \ ,
\end{equation}
where $A$ is a positive constant, resulting from the ion-ion interactions; and  $B$ is related to the ion-solvent interactions.
For kosmotropes
$B$-coefficient is positive, while for chaotropes it is negative, see \ref{tab2}.  In particular, we observe that \ce{F-} is a kosmotrope, while \ce{Br-} and \ce{I-}  are chaotropes. Chloride ion appears to be on the border line between the two regimes.
The classification of ions into kosmotropes and chaotropes correlates well with our theory of surface tensions of halogen salts  --- kosmotropes were found to remain strongly hydrated near the GDS, while chaotropes lost their hydration sheath near the interface. 
It is curious that the bulk dynamics of ion-water interaction, measured by the value of the viscosity $B$-coefficient, is so strongly correlated with the statics of ionic hydration, measured by the surface tension of electrolyte solutions.

\begin{table}
   \centering
   \setlength{\belowcaptionskip}{10pt}
   \caption{$B$-coefficients for various ions \cite{JeMa95}.}
   \begin{tabular}{|c|c|c|c|}
      \hline
      Ions            & $B$-coefficient & Ions            & $B$-coefficient \\
      \hline
      \ce{Na+}        & 0.085           & \ce{BrO3-}      & 0.009  \\
      \hline
      \ce{F-}         & 0.107           & \ce{NO3-}       & -0.043 \\
      \hline
      \ce{Cl-}        & -0.005          & \ce{ClO3-}      & -0.022 \\
      \hline
      \ce{Br-}        & -0.033          & \ce{ClO4-}      & -0.058 \\
      \hline
      \ce{I-}         & -0.073          & \ce{CO3^2-}     & 0.294  \\
      \hline
      \ce{IO3-}       & 0.140           & \ce{SO4^2-}     & 0.206  \\
      \hline
   \end{tabular}
   \label{tab2}
\end{table}

It is reasonable, then,  to suppose that the classification of ions into chaotropes and kosmotropes,  based on their $B$-coefficient, will also 
extend to more complicated ions as well.  Thus, we expect that chaotropes \ce{NO3-}, \ce{ClO3-} and \ce{ClO4-}, will loose their hydration sheath
near the GDS, while the kosmotropes \ce{IO3-}, \ce{CO3^2-} and \ce{SO4^2-},  will remain strongly hydrated.  Furthermore, since $B$-coefficients
of these ions are large (as compared to  \ce{F-}), we expect that these ions will remain as hydrated near the GDS as they 
were in the bulk, similar to what  was found for fluoride anion.  On the other hand, the $B$-coefficient of
\ce{BrO3-} is very close to zero and, similarly  to  \ce{Cl-}, we expect bromate  to be only very weakly hydrated near the GDS.

There is an additional difficulty with studying oxy-anions. Since these ions are not spherically symmetric, their 
radius is not well defined.  Nevertheless, it has been observed that empirical radius
\begin{equation}
a_0 = \frac{ n_{oxy}}{4} \left(d+1.4 \text{\ \AA}\right)
\label{size}
\end{equation}
where $d$ is the \ce{M}-\ce{O} covalent bond length in the corresponding salt crystal and $n_{oxy}$  is the number of oxygens in anion,
correlates very well with the experimental measurements of entropies of hydration \cite{CoLa57}.  Using this formula
we calculate: $a_{NO_3}=1.98$ \AA, $a_{ClO_3}=2.16$ \AA\ and $a_{ClO_4}=2.83$ \AA, for the bare radius of the chaotropic oxy-anions.
The polarizabilities of \ce{NO3-} and  \ce{ClO4-}  are given in the reference \cite{PyPi92}: $\gamma_{NO_3}=4.09$ \AA$^3$ and $\gamma_{ClO_4}=5.4$ \AA$^3$. Unfortunately, this reference does not provide the polarizability of chlorate ion.
\begin{figure}
\begin{center}
\includegraphics[width=7.29cm]{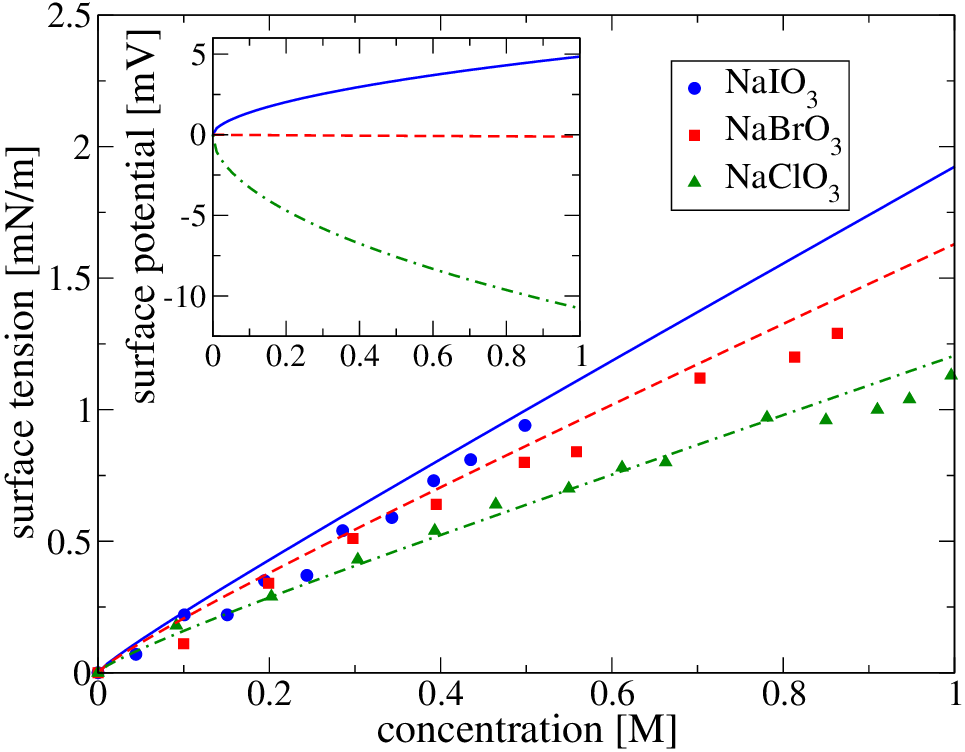}
\end{center}
\caption{Surface tensions for \ce{NaIO3}, \ce{NaBrO3} and \ce{NaClO3}. The symbols are experimental data \cite{Ma} and the lines are obtained using the present theory. The inset shows the electrostatic surface potential difference and has the same horizontal axis label as the main figure.}
\label{fig4}
\end{figure}
However,
since the polarizability is, in general, proportional to the volume of an ion, we can easily estimate it  based
on the polarizability of iodate, which is given in the reference \cite{PyPi92}, $\gamma_{IO_3}=8.0$ \AA$^3$.
We then obtain  $\gamma_{ClO_3}=5.3$ \AA$^3$.
Using eq \ref{e7}, we can now calculate the surface potential, the ionic density distribution and, integrating the Gibbs adsorption isotherm, the excess surface tension, \ref{fig4}, \ref{fig5}, \ref{fig6}, \ref{fig7}, and \ref{fig8}. We stress that in all the calculations we are using the same
value of sodium hydrated radius,  $a_{h}=2.5$ \AA, obtained for halogen salts.   For nitrate and chlorate, the  predictions of the theory
are in good agreement with the experimental
measurements of Matubayasi \cite{Ma}, see \ref{fig4} and \ref{fig5}.

\begin{figure}
\begin{center}
\includegraphics[width=7.29cm]{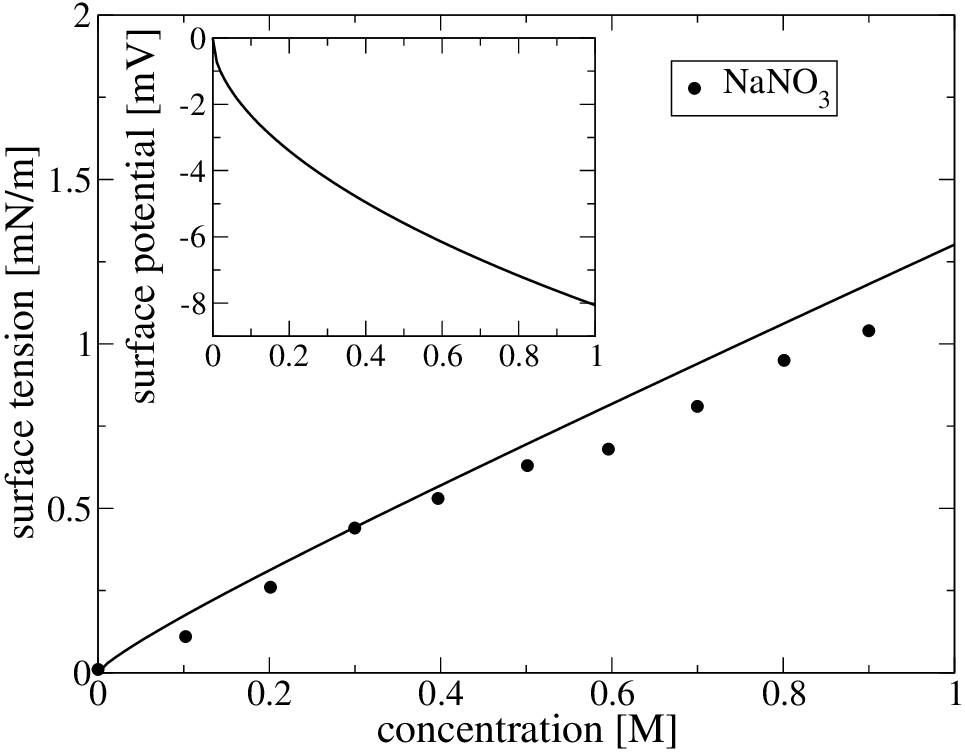}
\end{center}
\caption{Surface tension of \ce{NaNO3}. The symbols are the experimental data \cite{Ma} and the lines are the present theory. 
The inset shows the electrostatic surface potential difference vs. the molar concentration.}
\label{fig5}
\end{figure}
\begin{figure}
\begin{center}
\includegraphics[width=7.29cm]{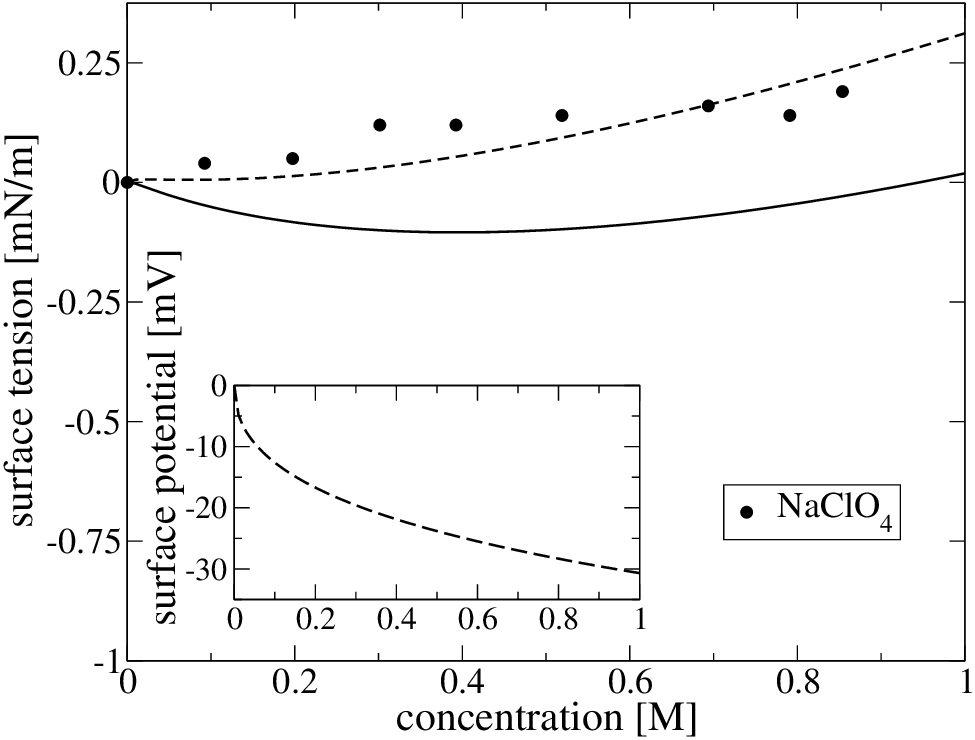}
\end{center}
\caption{Surface tensions of \ce{NaClO4}. The symbols are the experimental data \cite{Ma} and the lines are the present theory. The solid line is obtained using $a_{ClO_4}=2.83$ \AA\ from eq. (\ref{size}), and the dashed line using $a_{ClO_4}=2.75$ \AA. The inset shows the electrostatic surface potential difference vs. the molar concentration.}
\label{fig6}
\end{figure}
For perchlorate, the theoretical surface tension is slightly lower than what was found
experimentally, \ref{fig6}. The difficulty is that since the cavitational energy scales with the volume of the ion, for large ions it becomes very sensitive to the precise value of the ionic radius. For example, if we use a slightly smaller radius of perchlorate  $a_{ClO_4}=2.75$ \AA, we obtain a good
agreement with experiment, \ref{fig6}.
The theory shows that \ce{ClO4-} is adsorbed stronger than either \ce{ClO3-} or \ce{I-}, see \ref{fig7} and \ref{fig8}. This is conclusion is also in agreement with the simulations \cite{OtVa09}.
\begin{figure}
\begin{center}
\includegraphics[width=7.29cm]{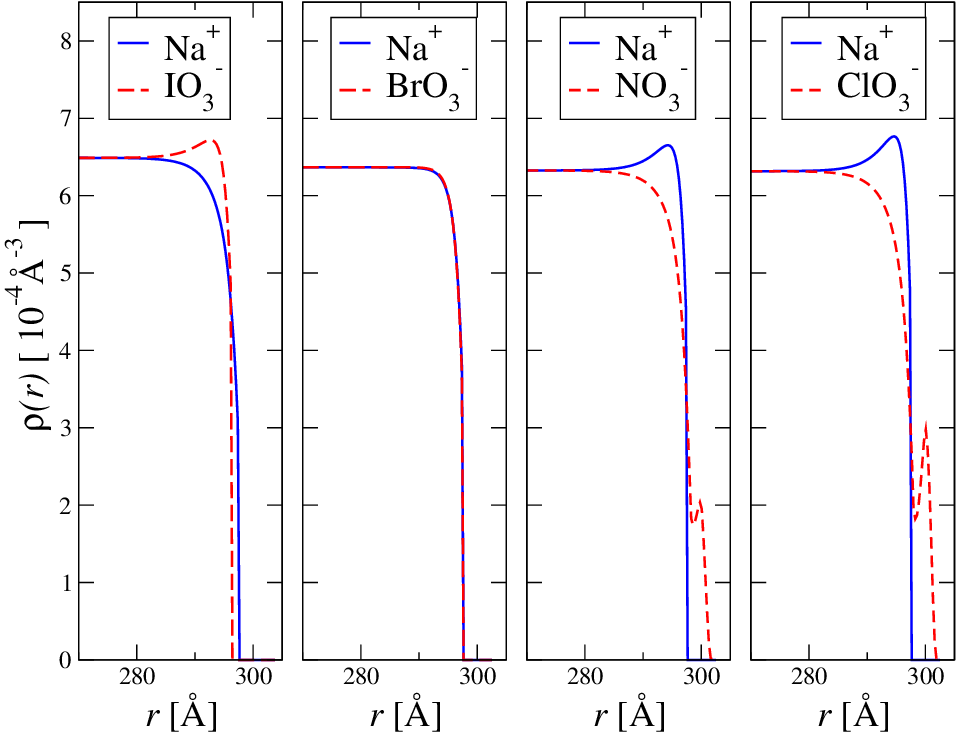}
\end{center}
\caption{Theoretical ionic density profiles for \ce{NaIO3}, \ce{NaBrO3}, \ce{NaNO3} and \ce{NaClO3} at $1$M. The GDS is at $r=300$ \AA.}
\label{fig7}
\end{figure}
\begin{figure}
\begin{center}
\includegraphics[width=7.29cm]{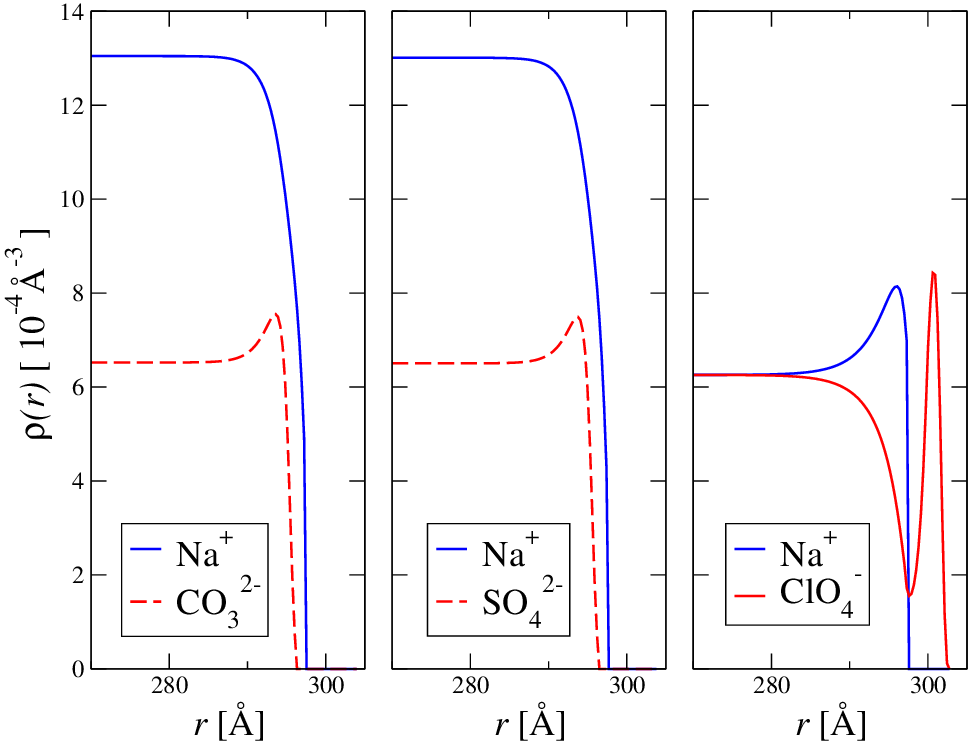}
\end{center}
\caption{Theoretical ionic density profiles for \ce{Na2CO3}, \ce{Na2SO4} and \ce{NaClO4} at $1$M. The GDS is at $r=300$ \AA.}
\label{fig8}
\end{figure}

We next calculate the surface tension of sodium salts with kosmotropic anions.
Since the $B$-coefficient of iodate is so large, we expect that this ion
will remain completely hydrated  near the GDS.  The bulk hydrated radius  of  \ce{IO3-}  is $a_h=3.74$ \AA \cite{Ni59}.  Using this
in eq \ref{e7a}, we obtain an excellent agreement with the experimental data \cite{Ma}, \ref{fig4}.  The  $B$-coefficient of bromate is almost zero,
so that \ce{BrO3-}  should be only very weakly hydrated near the GDS.
Indeed, using $a_h=2.41$ \AA, we obtain a very good agreement with the experimental data \cite{Ma}, \ref{fig4}.
This value is only slightly above the crystallographic radius of \ce{BrO3-}, $a_{BrO_3}=2.31$ \AA, obtained using eq \ref{size}.
\begin{figure}
\begin{center}
\includegraphics[width=7.29cm]{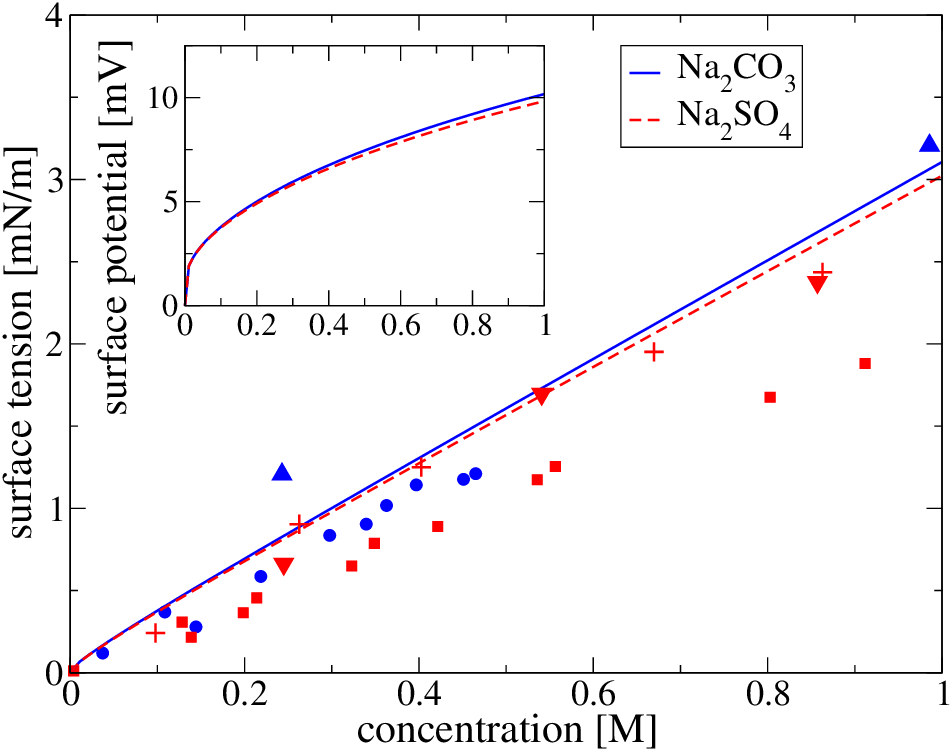}
\end{center}
\caption{Surface tensions of \ce{Na2CO3} and \ce{Na2SO4}. The circles and the squares are the experimental data for sodium carbonate and soldium sulfate, respectively \cite{MaTs01} . The up and down triangles are the data of Jarvis  and Scheiman \cite{JaSc68}. The plus symbols are the data of Weissenborn and Pugh \cite{WePu96} for the \ce{Na2SO4} salt. The lines are the present theory. The inset shows the surface potential difference vs. the molar concentration.}
\label{fig9}
\end{figure}

Finally, we consider the surface tension of sodium salts with divalent anion. The predictions of the theory, in this case, should be taken with a grain of salt, since it is well known that Poisson-Boltzmann theory begins to loose its validity for divalent ion as a result of inter-ionic correlations \cite{Le02}.  Nevertheless, it is curious to see how well the theory does in this limiting case.
The $B$-coefficients of \ce{SO4^2-} and \ce{CO3^2-} are large and positive, so that we expect these ions to be fully hydrated, $a_h=3.79$ \AA\ and $3.94$ \AA\ \cite{Ni59}, respectively.  Calculating the surface tensions, we find that our results are somewhat above 
those measured by  Matubayasi et al.  \cite{MaTs01} .  Curiously, the theory agrees well
the measurements of Jarvis and Scheiman \cite{JaSc68} and of Weissenborn and Pugh \cite{WePu96}, \ref{fig9}.  At this point we do not  know what to make of this agreement.  The surface potentials of sodium sulfate and sodium carbonate are listed in \ref{tab1}. In \ref{tab3} we summarize the classification of all ions and the radii used in calculations.
\begin{table}
\centering
\setlength{\belowcaptionskip}{10pt}
\caption{Ion classification into chaotropes (c) and  kosmotropes (k) and their effective radii. The kosmotropic radii listed are either hydrated or partially hydrated. The chaotropic radii are the bare ones,  for oxy-anions  given by eq \ref{size}.}
      \begin{tabular}{|c|c|c|}
      \hline
      Ions            &    chao/kosmo     & radius (\AA) \\
      \hline
      \ce{Na+}        &        k          &       2.5    \\
      \hline
      \ce{F-}         &        k          &       3.54   \\
      \hline
      \ce{Cl-}        &        k          &       2      \\
      \hline
      \ce{Br-}        &        c          &       2.05   \\
      \hline
      \ce{I-}         &        c          &       2.26   \\
      \hline
      \ce{IO3-}       &        k          &       3.74   \\
      \hline
      \ce{BrO3-}      &        k          &       2.41   \\
      \hline
      \ce{NO3-}       &        c          &       1.98   \\
      \hline
      \ce{ClO3-}      &        c          &       2.16   \\
      \hline
      \ce{ClO4-}      &        c          &       2.83   \\
      \hline
      \ce{CO3^2-}     &        k          &       3.94   \\
      \hline
      \ce{SO4^2-}     &        k          &       3.79   \\
      \hline
   \end{tabular}
   \label{tab3}
\end{table}

\section{Conclusions}

We have presented a theory which allows us to quantitatively calculate the surface tensions and the surface potentials of
10 sodium salts with only one adjustable parameter, the partial hydrated radius of the sodium cation.  We find that anions near
the GDS can be classified into kosmotropes and chaotropes.  Kosmotropes remain hydrated near the interface, while the
chaotropes loose their hydration sheath.  For all salts studied in this paper, the classification of ions into the structure makers/breakers
agrees with the bulk characterization based on the Jones-Dole viscosity $B$-coefficient.   If anions
are arranged in the order of increasing electrostatic surface potential difference, we obtain the extended lyotropic (Hofmeister) series,
\ce{CO3^2-} > \ce{SO4^2-} > \ce{IO3-} > \ce{F-} > \ce{BrO3-} > \ce{Cl-} > \ce{NO3-} > \ce{Br-} > \ce{ClO3-} > \ce{I-} > \ce{ClO4-}.
To our knowledge this is the first time that this sequence has been derived theoretically.  The theory also helps to understand why
Hofmeister series is relevant to biology \cite{CoWa85,*ZhCr06}. In general, proteins have both hydrophobic and hydrophilic moieties.  If the hydrophobic region
is sufficiently large, it can become completely dewetted \cite{LuCh99,*Ch05},  surrounded by a vapor-like film into which chaotropic anions
can become adsorbed.  This will help to solvate proteins, but at the same time will destroy their tertiary structure, 
denaturing them in the process.  On the other hand, the kosmotropic ions will favor protein precipitation.  The action is again
two fold. The kosmotropes  adsorb  water  which could,  otherwise, be used to solvate hydrophilic moieties.
They also screen the electrostatic repulsion, allowing charged proteins to come into 
a close contact and to stick together as the result of their van der Waals and hydrophobic interactions.  
This  leads to formation of large clusters which can no longer be  dispersed in water and must precipitate.

\begin{acknowledgement}

We thank Prof. Norihiro Matubayasi for providing us with the unpublished data for the surface tensions of
chlorate, bromate, iodate and perchlorate.  YL is also grateful to Dr. Delfi Bastos Gonz\'alez for first
bringing the kosmotropic nature of iodate to his attention.
\end{acknowledgement}

\bibliography{ref.bib}

\end{document}